# Motion correction in cardiac perfusion data by using robust matrix decomposition

"Abdul Haseeb Ahmed, Ijaz M. Qureshi" *

"*1*Air University, E-9, Islamabad and Pakistan"

**Abstract**

Motion free reconstruction of compressively sampled cardiac perfusion MR images is a challenging problem. It is due to the aliasing artifacts and the rapid contrast changes in the reconstructed perfusion images. In addition to the reconstruction limitations, many registration algorithms underperform in the presence of the rapid intensity changes. In this paper, we propose a novel motion correction method that reconstructs the motion free image series from the undersampled cardiac perfusion MR data. The motion correction method uses the novel robust principal component analysis based reconstruction along with the periodic decomposition to separate the respiratory motion component that can be registered, from the contrast intensity variations. It is tested on simulated data and the clinically acquired data. The performance of the method is qualitatively assessed and compared with the existing motion correction methods. The proposed method is validated by comparing manually acquired time-intensity curves of the myocardial sectors to automatically generated curves before and after registration.

## 1. Introduction

Contrast enhanced perfusion magnetic resonance imaging (MRI) is a reliable tool for the assessment of blood flow in the myocardium. MRI is used for the diagnosis of coronary artery disease, which leads to reduced blood supply to the myocardium [1] [2]. Perfusion image series are acquired to cover pre-contrast baseline images and the full cycle of contrast flow, which span over 40-60 sec generally. First, contrast agent enters the right ventricle (RV), then the left ventricle (LV), and finally, the agent flow into the LV myocardium. Then, to measure the blood flow, the image intensity of areas in the myocardium is tracked over time. The automatic assessment of the intensity change over time requires no motion artifacts during image acquisition. The moving heart is always imaged at the same cardiac phase, which is achieved by ECG based synchronization. The 60 sec acquisition time span is long for average patients to hold their breath, which leads to the breathing artifacts in the image series. An added challenge to motion correction is posed by the contrast agent passing through the heart that results in a strong intensity change over time [3-5].

* Corresponding author:
E-mail address:haseeb@mail.au.edu.pk

There are two major problems with cardiac perfusion MRI:

- Respiratory motion artifacts
- Strong intensity change that is induced by the contrast agent.

Several image registration methods were proposed to overcome the motion artifacts. However, the registration based methods were not able to distinguish between the motion artifacts and the local intensity changes across different time points [6]. Therefore, some techniques were based on the separation of respiratory motion and contrast enhancement before image registration. The independent component analysis based method was proposed to create the synthetic images that were used as reference for linear image registration [7]. Due to complex deformations of myocardium, non-linear registration methods were proposed to overcome the limitations of linear registration [8, 9]. Melbourne et al. proposed progressive principal component analysis, that was based on the iterative use of principal component analysis and image registration, which gradually eradicated the misalignments. However, this method unable to separate the dominant contrast agent component and the respiratory motion component [10]. Wollny et al. employed the independent component analysis to decompose the perfusion images before registration but failed to separate the respiratory motion from the contrast changes [11]. Recently, Hamy et al. proposed the respiratory motion correction method based on the robust principal component analysis, to decompose the dynamic MRI into a low rank (respiratory motion) and a sparse component (contrast changes). However, this method failed to remove the respiratory motion from the contrast agent [12].

In the recent past, Compressed Sensing (CS) has emerged as a useful tool to overcome the image acquisition limitations in MRI. CS exploits the MR image sparsity and incoherence between the acquisition space and the representation space to reconstruct undersampled data without compromising quality [13-15]. CS based reconstruction is done by enforcing sparsity in the solution, subject to data consistency constraints. CS is well suited for the dynamic images, because they provide sparser representations in the spatiotemporal domain as compared to the spatial domain alone.

Recently, RPCA was proposed to decompose the data matrix as a superposition of a low rank matrix (L) and sparse matrix (S) [16]. RPCA addressed the shortcomings of classical PCA in the presence of grossly corrupted observations, in which low rank component and sparse component were separated completely by convex programing. The combination of compressed sensing and robust principal component analysis were exploited to increase the image acquisition speed with the separation of back ground and dynamic components [17]. RPCA was well suited for the dynamic contrast enhanced imaging, as L modelled the temporally correlated background and S captured the dynamic information.

In this paper, a new motion correction method is proposed to address the problem of respiratory motion in the compressively sampled cardiac perfusion MRI. We propose a new RPCA based reconstruction method to separate the low rank component, the sparse component and the acquisition noise. After image reconstruction from the compressively sampled data, the method separates the respiratory motion from the contrast motion using periodic decomposition. Finally, a non-rigid registration is used to correct the respiratory motion artifacts in the cardiac perfusion MRI. The performance of our method is stage wise evaluated using simulated and in vivo data. We also perform a comparison with the existing methods to show the improvements in the reconstructed MRI.

## 2. Theory

Cardiac perfusion images can be efficiently represented as a superposition of low rank component, which contains smooth and slowly varying changes (some respiratory motion), and sparse component, which captures the fast and local intensity changes (contrast enhancement). RPCA is aimed to decompose the perfusion images as a superposition of motion related changes, which are periodic in nature, and local changes caused by contrast enhancement, which are non-periodic. Since respiratory motion (periodic component) and contrast changes (non-periodic) cannot be perfectly separated with RPCA, therefore, a balancing parameter is used to control the amount of motion included in the low rank component [1, 2]. With this control, it is possible to obtain the low rank component with no motion and the sparse component with both periodic and non-periodic motions.

The reconstruction of undersampled perfusion data requires three types of incoherence as stated by the Otazo et al. [3]. The first two incoherence are related to the acquisition space ($k$-$t$) and the representation space ($L$, $S$), which are required to remove aliasing artifacts. The third incoherence is between the low rank component and the sparse component.

Since the acquisition of k-space samples are often corrupted by noise which effects the reconstructed images. Therefore, the small entry wise noise is added to the RPCA model to make the low rank and sparse decomposition stable [4].

*2.1. First stage*

Let matrix $M \in \mathbb{R}^{n \times t}$ be a Casorati matrix with each column being formed by each frame of perfusion image series. RPCA decomposes $M$ as a superposition of a low-rank matrix $L$ with few non-zero singular values and a sparse matrix $S$ with few non-zero entries (where $n = n1 \times n2$). RPCA is a well posed problem with a unique solution if the incoherence mentioned above exist. RPCA was modified to reconstruct the undersampled and noisy perfusion MRI data by solving the following convex optimization problem [3] [4]:

$$\min_{L,S} \lambda_L \|L\|_* + \lambda_S \|S\|_1$$
$$\text{subject to} \quad \|F_u(L+S)-D\|_F^2 \leq \delta \tag{1}$$

Where,

- $F_u$ is undersampled spatial Fourier operator,
- $\|L\|_*$ is the nuclear norm which is calculated as the sum of singular values of the matrix $L$,
- $\|S\|_1$ is the $l_1$-norm which is calculated by sum of absolute values of the entries of $S$.
- $\|\cdot\|_F$ is the Frobenius norm.
- $D = F_u M$ is the undersampled noisy k space data.
- $\lambda_L$ and $\lambda_S$ are the tradeoff parameters.

The Equation 1 is reformulated as follow [4, 5]:

$$\min_{L,S,Z} \lambda_L \|L\|_* + \lambda_S \|S\|_1 + \frac{1}{2\mu}\|F_u Z\|_F^2$$
$$\text{subject to} \quad L+S+Z = M \tag{2}$$

Where,

- $Z$ is a small entry wise noise term.
- $\mu$ is the smallest value to threshold away the noise, but not too large to shrink the $L$ and $S$.

In order to minimize the Equation 2, the augmented lagrangian function is written as:

$$\min_{L,S,Z} \lambda_L \|L\|_* + \lambda_S \|S\|_1 + \frac{1}{2\mu}\|F_u Z\|_F^2 - \langle Y, L+S+Z-M \rangle + \frac{\beta}{2}\|L+S+Z-M\|_F^2 \tag{3}$$

Where,

- $Y$ is the lagrangian multiplier.
- $\langle \cdot \rangle$ is the standard trace inner product.
- $\beta > 0$ is the penalty parameter for the violation of the linear constraint.

The augmented lagrangian function is solved by the Alternating direction multiplier method (ADMM) which minimizes the variables $L$, $S$ and $Z$, in Equation 3, independently.

$$L_{k+1} = \mathbb{S}_{\lambda_L/\beta}\left(\frac{Y_k}{\beta} + M - S_k - Z_k\right) = U_k \, shrink\left(\Sigma_k, \frac{\lambda_L}{\beta}\right) V_k^*$$

$$S_{k+1} = shrink\left(\frac{Y_k}{\beta} + M - L_{k+1} - Z_k, \frac{\lambda_S}{\beta}\right)$$

$$Z_{k+1} = \left(\frac{Y_k}{\beta} + M - L_{k+1} - S_{k+1}\right)\left(\frac{\mu\beta}{1+\mu\beta}\right)$$

$$Y_{k+1} = Y_k - \beta(L_{k+1} + S_{k+1} + Z_{k+1} - M)$$

(4)

In order to solve each sub-problems in parallel, Jacobian decomposition is applied to ADMM scheme [6, 7]. This feature is useful in big data scenario and reduces the computation time at each iteration. It is formulated as:

$$L_{k+1} = \mathbb{S}_{\lambda_L/\beta}\left(\frac{Y_k}{\beta} + M - S_k - Z_k\right) = U_k \, shrink\left(\Sigma_k, \frac{\lambda_L}{\beta}\right) V_k^*$$

$$S_{k+1} = shrink\left(\frac{Y_k}{\beta} + M - L_k - Z_k, \frac{\lambda_S}{\beta}\right)$$

$$Z_{k+1} = \left(\frac{Y_k}{\beta} + M - L_k - S_k\right)\left(\frac{\mu\beta}{1+\mu\beta}\right)$$

$$Y_{k+1} = Y_k - \beta(L_{k+1} + S_{k+1} + Z_{k+1} - M)$$

(5)

Equation 5 is used to solve each variable in parallel to reduce the computation time at each iteration. However, the convergence of the Equation 5 cannot be confirmed. It is proved that the convergence is ensured if the variables to be minimized in Equation 5, are regularized by quadratic proximal terms [7]. Equation 5 is modified as:

$$L_{k+1} = \mathbb{S}_{\lambda_L/\beta}\left(\frac{Y_k}{\beta} + M + \tau L_k - S_k - Z_k\right) = U_k \, shrink\left(\Sigma_k, \frac{\lambda_L}{\beta}\right) V_k^*$$

$$S_{k+1} = shrink\left(\frac{Y_k}{\beta} + M + \tau S_k - L_k - Z_k, \frac{\lambda_S}{\beta}\right)$$

$$Z_{k+1} = \left(\frac{Y_k}{\beta} + M + \tau Z_k - L_k - S_k\right)\left(\frac{\mu\beta}{1+\mu\beta}\right)$$

$$Y_{k+1} = Y_k - \beta(L_{k+1} + S_{k+1} + Z_{k+1} - M)$$

(6)

Since it is not possible to separate the respiratory motion and the contrast changes completely, therefore, a stage wise approach is proposed in which the first stage gives the low rank component *L*, which contains the background image series with no motion and the sparse component *S*, which is the combination of respiratory motion and the motion due to contrast changes.

## 2.2. Second stage

The output of the first stage, low rank component $L$ and sparse component $S$, are fed to the second stage which separates the periodic motion and the non-periodic motion iteratively. The sparse component is decomposed into the periodic component and the non-periodic component using the periodicity decomposition based method. Each row of the sparse component $S$ contains the periodicity, which is separated from the non-periodic component using periodic decomposition method [8].

Let $I_p$ is the identity matrix of dimension $p \times p$ with the period p and $C_p$ is the periodic matrix with dimension $t \times t$ defines as:

$$C_p = \begin{bmatrix} Ip & \cdots & Ip \\ \vdots & \ddots & \vdots \\ Ip & \cdots & Ip \end{bmatrix}$$

To extract the periodic component from the sparse matrix S, an equation is defined as:

$$P = S \cdot C_p$$

Where $P$ is the periodic component. In order to find the matrix $C_p$ that searches for the best periodic characterization of the given image series, m-best method is used as proposed by the Sethares et.al. [8].

This stage decomposes the $S$ into the periodic component $P$ (respiratory motion) and the non-periodic component $Q$ (contrast motion) as mentioned in the Equation 7 given below:

$$S = P + Q \tag{7}$$

## 2.3. Third stage

In the third stage, we have the low rank component $L$ (back ground with no motion), periodic motion image series $P$ and the non-periodic component $Q$ (only contains motion due to contrast agent). The first two components are combined to give the low rank component $L_p$ with still background and the respiratory motion (periodic component). A demon based non-rigid registration technique is used to register the low rank frames in order to remove the respiratory motion from the low rank component [9]. Since the registration technique is computational simple and robust to the intensity changes, therefore, it is used to correct the respiratory motion artifacts. Demon based registration is given as:

$$U_d = \frac{(x_d - x_1)\nabla x_1}{|\nabla x_1|^2 + \alpha(x_d - x_1)^2} + \frac{(x_d - x_1)\nabla x_d}{|\nabla x_d|^2 + \alpha(x_d - x_1)^2} \tag{8}$$

After the registration process $\Gamma$, the registered low rank $L_{reg}$ is obtained by using following Equation 9:

$$L_{reg} = \Gamma(L_p)$$

The registered low rank $L_{reg}$ and the non-periodic component $Q$ are combined to give the cardiac perfusion image series with respiratory motion correction $M_{mc}$ as given below:

$$M_{mc} = L_{reg} + Q \qquad (9)$$

## 3. Materials and Methods

*3.1. Simulated data*

A framework for realistic cardiac MR simulations was proposed to evaluate the performance analysis of the image reconstruction algorithms from undersampled data in the presence of motion [10]. Free breathing myocardial perfusion phantoms were generated with image dimension=224x192x32 and pixel resolution =2mm x 2mm x 5mm. Periodic motion decomposition was performed for the base level slice. The purpose of this data is to show the performance of the proposed technique, as the technique performs better in the simulated data as compared to the real data.

*3.2. Clinical data*

Wollny et al has provided first-pass contrast-enhanced myocardial perfusion imaging data sets, which were acquired and processed under clinical research protocols and all subjects provided written informed consent [11]. Two distinct pulse sequences were employed for image acquisition: A hybrid GRE-EPI sequence and a true-FISP sequence. Both sequences were ECG-triggered and used 90-degree-saturation recovery imaging of several slices per R-R interval acquired for 60 heartbeats. the true-FISP sequence parameters were 50-degree readout flip angle, 975 Hz/pixel bandwidth, TE/TR/TI = 1.3/2.8/90 ms, 128 x 88 matrix,6 mm slice thickness. The GRE-EPI sequence parameters were 25-degree readout flip angle, echo train length = 4, 1500 Hz/pixel bandwidth, TE/TR/TI = 1.1/6.5/70 ms, 128 x 96 matrix, 8 mm slice thickness. The spatial resolution was about 2.8 mm x 3.5 mm. For all patients, a half dose of contrast agent (Gd-DTPA, 0.1 mmol/kg) was administered at 2.5 ml/s, followed by saline flush. All data sets were acquired using a free breathing protocol. Motion correction was performed for base-level slice with 58 perfusion images per slice.

*3.3. Performance Analysis*

Time–intensity curves of the myocardium are the prominent feature based on which the medical investigations can be done. Therefore, we have based our performance comparison on the time intensity curves for the six sectors defined in the American Heart Association (ASA) model of the myocardium. We compared the manually acquired time–intensity curves of 6 sections and the time intensity curves of the automatically obtained myocardium before registration and after registration. To estimate these time–intensity curves, for all data sets in base slice the LV-myocardium was segmented and the center of the LV as well as the middle of septum from the short-axis images was identified. Using the LV center and the middle of septum, the myocardium was segmented clock-wise into 6 segments enclosing equal angles. Finally, time–intensity curves were estimated based on the average pixel intensities of these segments.

## 4. Results

*4.1. First Stage*

The first stage image reconstruction is performed in MATLAB. The regularization parameters $\lambda_L$, $\lambda_S$ and $\mu$ are estimated empirically. The ranges are $\lambda_L$ = 0.002-0.01, $\lambda_S$ = 0.001-0.008 and $\mu$ = 0.2-2.5. The value of $\beta$= 0.5 and the value of $\tau$=0.001 are used in the results. The first stage reconstruction performance of simulated data and clinical data are evaluated using the root mean square error (RMSE). The reconstruction performance of the proposed method is compared with the *L+S method* [3]. For simulated perfusion images, undersampled Cartesian data is acquired at the rates *R= {2, 4, and 12}*. For clinical perfusion images, the fully sampled Cartesian data is retrospectively undersampled by the rates *R= {2, 4, 8, 12}* using the variable density sampling. If the acceleration rate *R* is further increased, the reconstructed image quality deteriorates rapidly [12]. For the comparison of the proposed method with the existing *L+S* method, the fully sampled perfusion image series are denoised using wavelets. Table 1 shows the performance of the proposed method in comparison with the *L+S* method for different acceleration rates using the simulated perfusion data. Table 2 shows the comparison of the proposed method with the *L+S* method for different acceleration rates using the in vivo perfusion rest data. Table 3 shows the performance of the proposed method in comparison with the *L+S* method for different acceleration rates using the in vivo perfusion stress data.

Table 1. Reconstruction Performance comparison: Proposed method and L+S method are compared using simulated data with different acceleration rates

| Acceleration rates | RMSE | |
|---|---|---|
| | L+S Method | Proposed Method |
| 2 | 2.6x10-3 | 2.5x10-3 |
| 4 | 2.7x10-3 | 2.6x10-3 |
| 12 | 3.9x10-3 | 3.0x10-3 |

Table 2. Reconstruction Performance comparison: Proposed method and L+S method are compared using in vivo perfusion rest data with different acceleration rates

| Acceleration rates | RMSE | |
|---|---|---|
| | L+S Method | Proposed Method |
| 2 | 1.1x10-2 | 7.0x10-3 |
| 4 | 1.8x10-2 | 1.3x10-2 |
| 8 | 3.2x10-2 | 2.2x10-2 |
| 12 | 3.9x10-2 | 2.8x10-2 |

Table 3. Reconstruction Performance comparison: Proposed method and L+S method are compared using in vivo perfusion stress data with different acceleration rates

| Acceleration rates | RMSE | |
|---|---|---|
| | L+S Method | Proposed Method |
| 2 | 1.1x10-2 | 9.0x10-3 |
| 4 | 2.7x10-2 | 2.0x10-2 |
| 8 | 4.4x10-2 | 3.1x10-2 |
| 12 | 4.8x10-2 | 3.9x10-2 |

## 4.2. Second Stage

The second stage is used to decompose the sparse component *S* into periodic P and non-periodic Q components. The stage decomposition performance is evaluated using the qualitative assessment of the periodic and the non-periodic images. The performance of the periodic decomposition is assessed using simulated perfusion data as shown in the Figure 1and the           Figure 2 with the acceleration rates

*R=2* and *R=12* respectively. Figure 3 and Figure 4 show the in vivo perfusion data with the acceleration rates *R=2* and *R=12* respectively. The first rows show the contrast images (L+Q) without respiratory motion and the second rows show the images with periodic motion (P). The second stage figures show the decomposition that separates the respiratory motion and contrast motion.

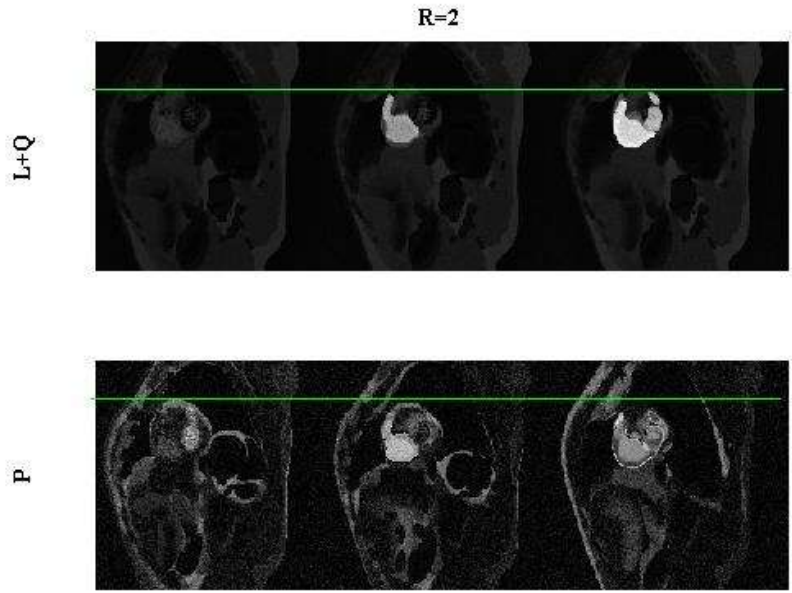

Figure 1: Decomposition of the periodic and non-periodic component of the simulated MRI with rate of 2. From top to bottom: sum of low rank component and contrast motion with three time indices of contrast flow; periodic component. Green line is used to indicate the respiratory motion in P and not present in L+Q.

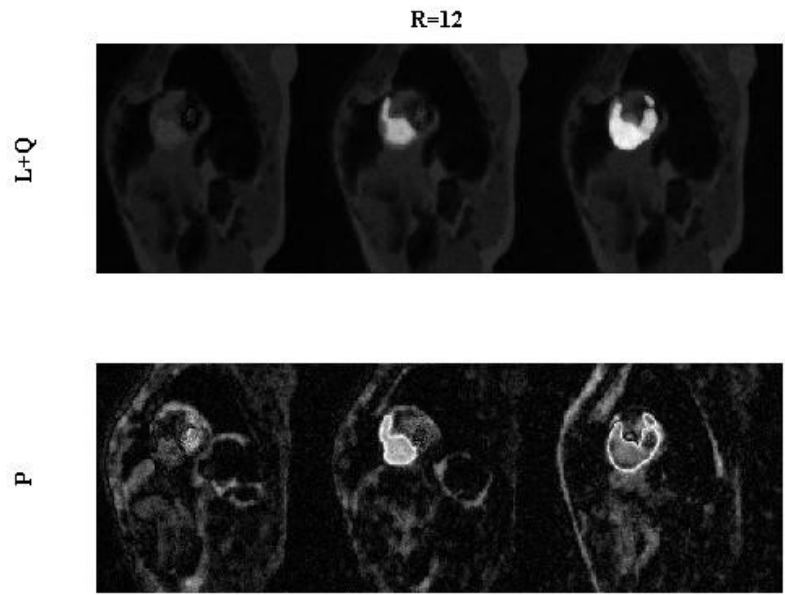

Figure 2: Decomposition of the periodic and non-periodic component of the simulated perfusion MRI with rate of 12. From top to bottom: sum of low rank component and contrast motion with three time indices of contrast flow; periodic component.

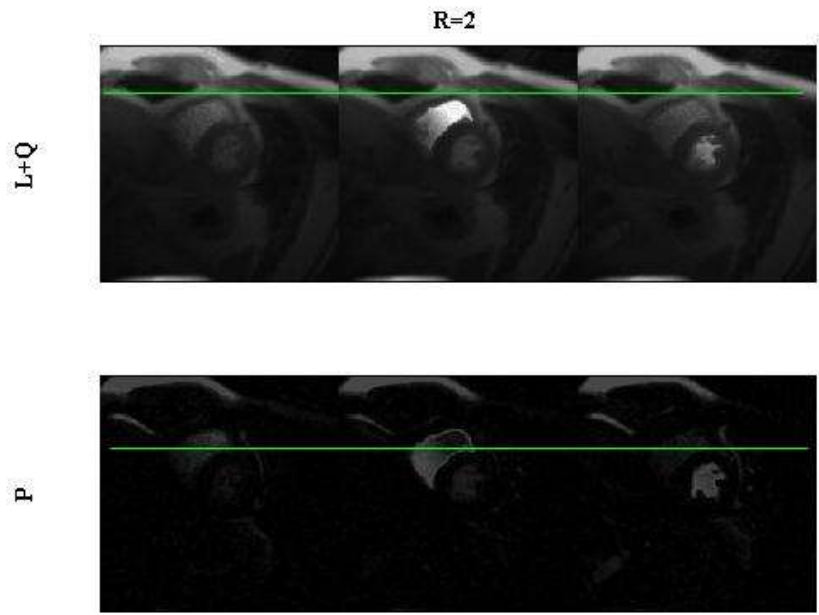

Figure 3: Decomposition of the periodic and non-periodic component of the in vivo perfusion MRI with rate of 2. From top to bottom: sum of low rank component and contrast motion with three time indices of contrast flow; periodic component. Green line is used to indicate the respiratory motion in P and not present in L+Q.

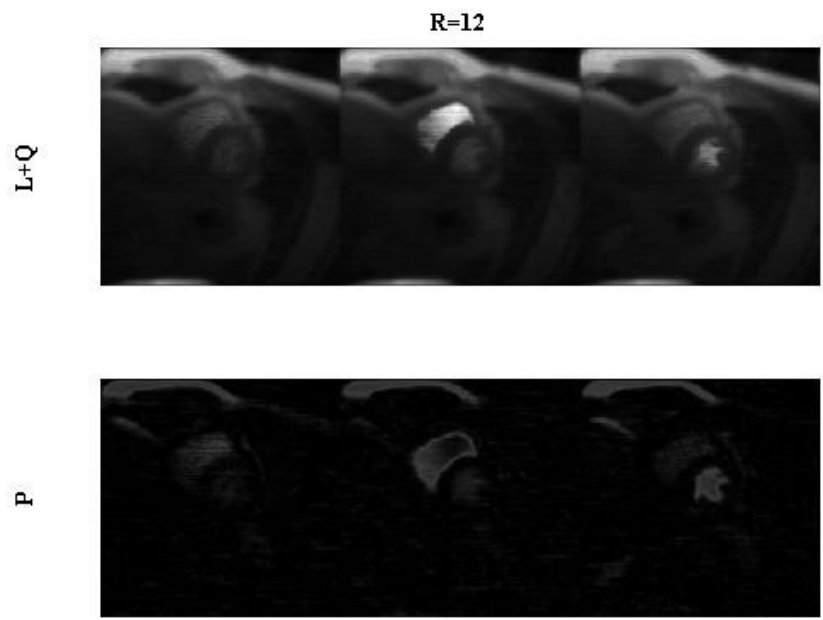

Figure 4: Decomposition of the periodic and non-periodic component of the perfusion MRI with rate of 2. From top to bottom: sum of low rank component and contrast motion with three time indices of contrast flow; periodic component. Green line is used to indicate the respiratory motion in P and not present in L+Q.

*4.3. Third Stage*

The implementation of the third stage is done in MATLAB according to the proposed method. The combination of the background *L* and the respiratory images *P* are used as input to the registration process. Figure 5 shows the result of the registered simulated data, using three acceleration rates R = {2, 4, and 12}. Figure 6 shows the result of the registered in vivo perfusion rest data, using four acceleration rates R = {2, 4, 8 and 12}. Figure 7 shows the result of the registered in vivo perfusion stress data, using four acceleration rates R = {2, 4, 8 and 12}. Four frames are used to show the result of the proposed method in the clinical images: pre-contrast, RV-peak, LV-peak and myocardial perfusion.

The output of the third stage is compared with the RPCA based motion correction method [2] in terms of the time intensity curve. Manually segmented time intensity curves are used as the reference curves. The time intensity curves of the proposed method followed the reference curves, whereas the other method curves illustrated erratic behavior, which is mainly because of the contrast motion with the respiratory motion and the lack of registration process.

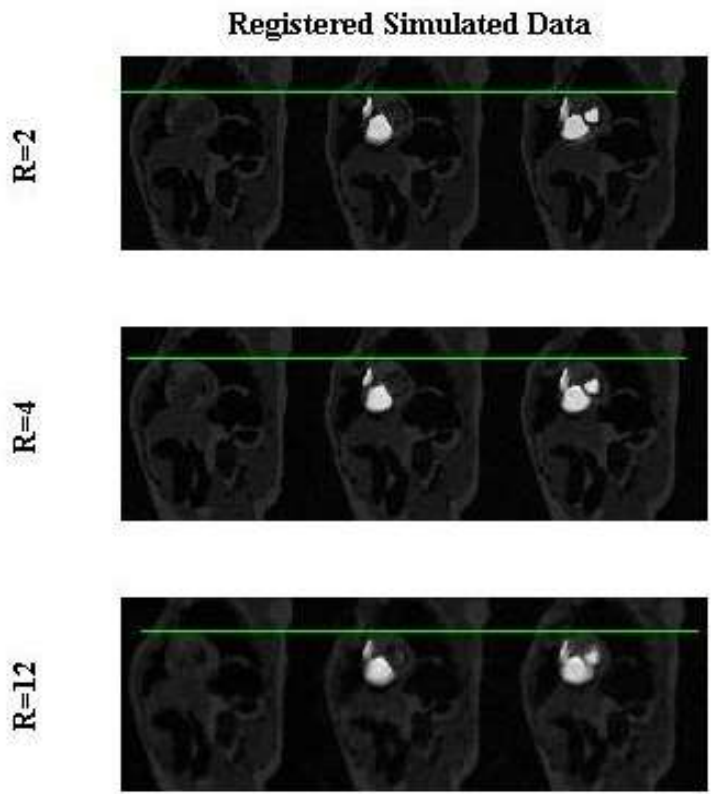

Figure 5: From top to bottom: Registered simulated perfusion MRI with acceleration rates of 2, 4 and 12. From top to bottom: Green line is used to indicate the presence of respiratory motion in three time indices of contrast flow in the heart.

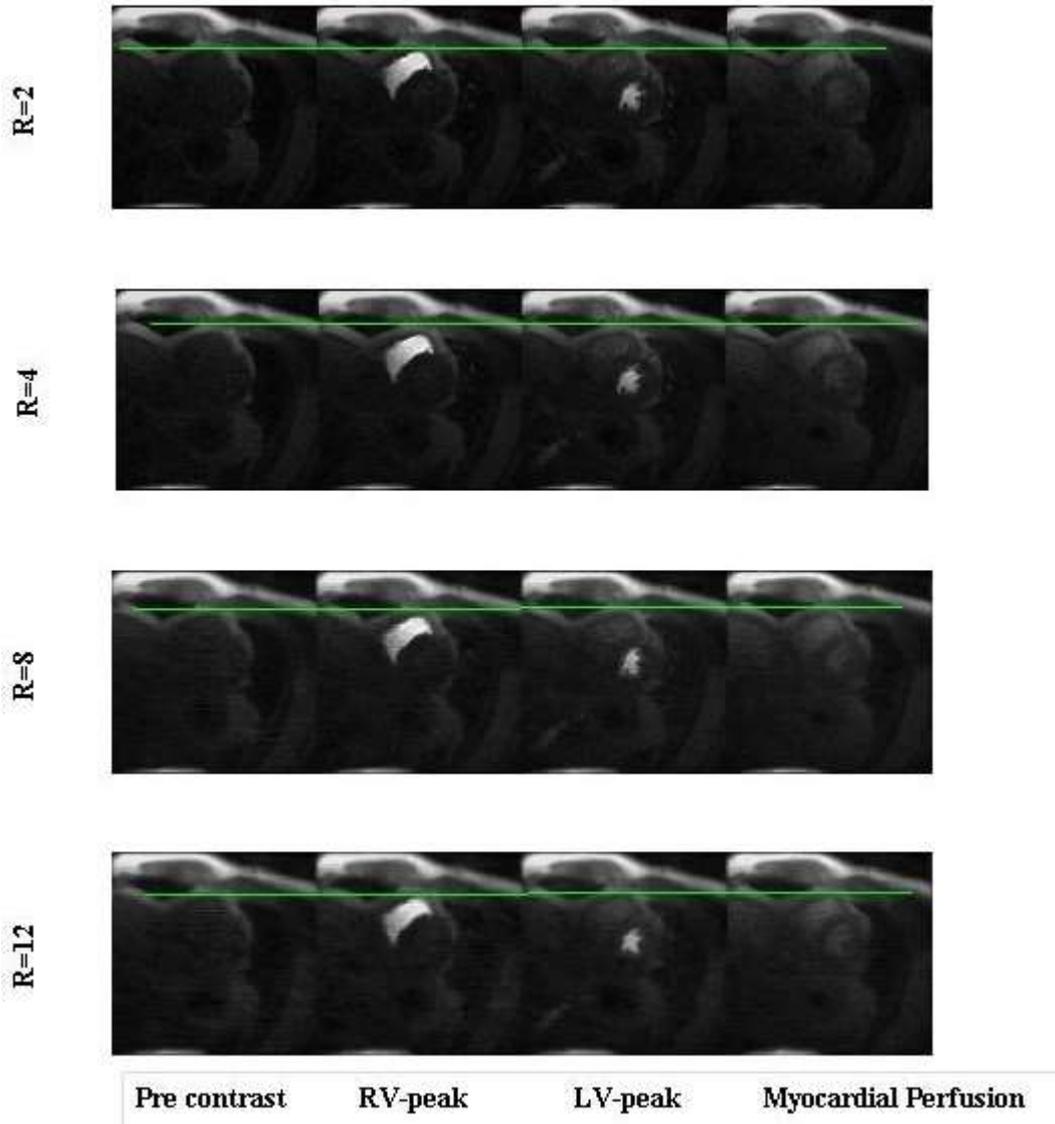

Figure 6: From top to bottom: Registered in vivo perfusion rest MRI with acceleration rates of 2, 4, 8 and 12. From top to bottom: Green line is used to indicate the presence of respiratory motion in four time indices of contrast flow in the heart; Pre-contrast, RV peak, LV peak and Myocardial perfusion.

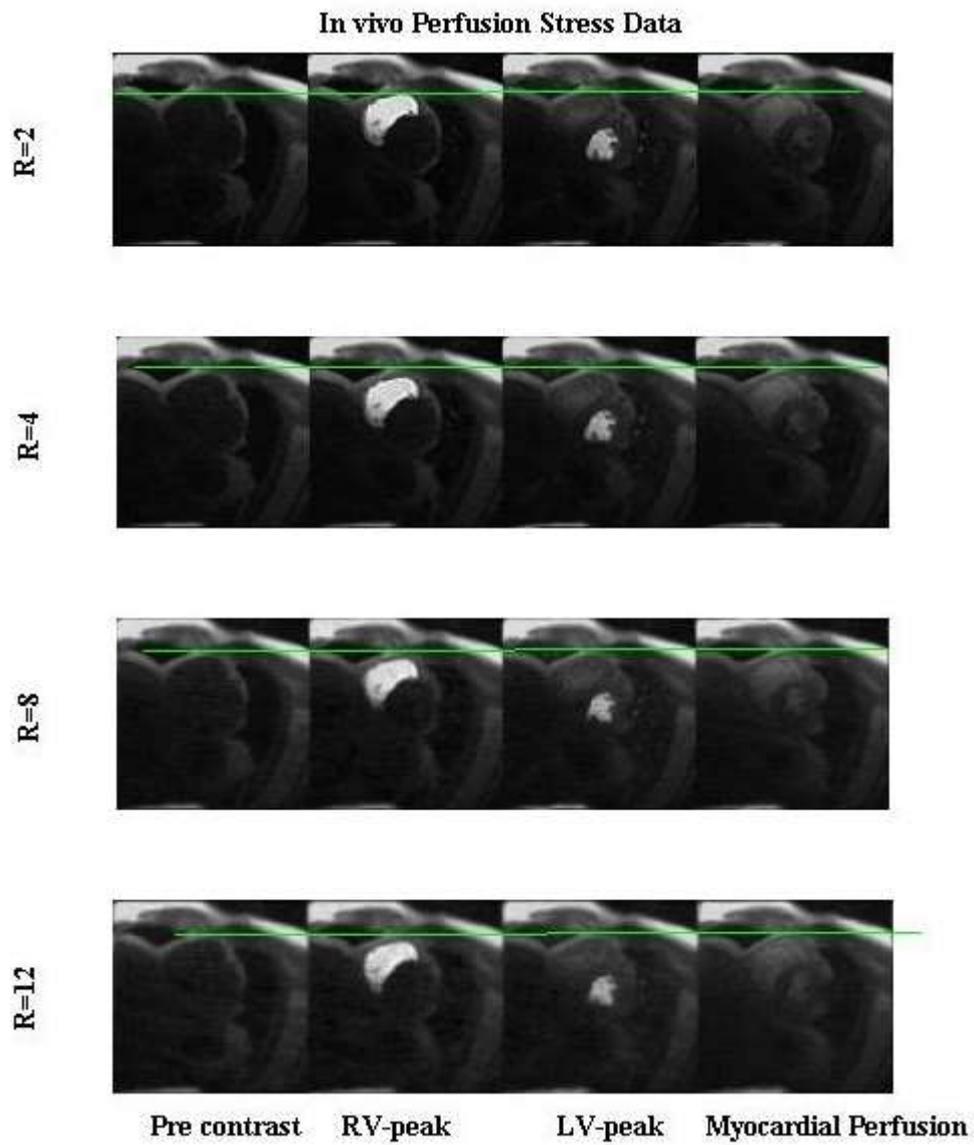

Figure 7: From top to bottom: Registered in vivo perfusion stress MRI with acceleration rates of 2, 4, 8 and 12. From top to bottom: Green line is used to indicate the presence of respiratory motion in four time indices of contrast flow in the heart; Pre-contrast, RV peak, LV peak and Myocardial perfusion.

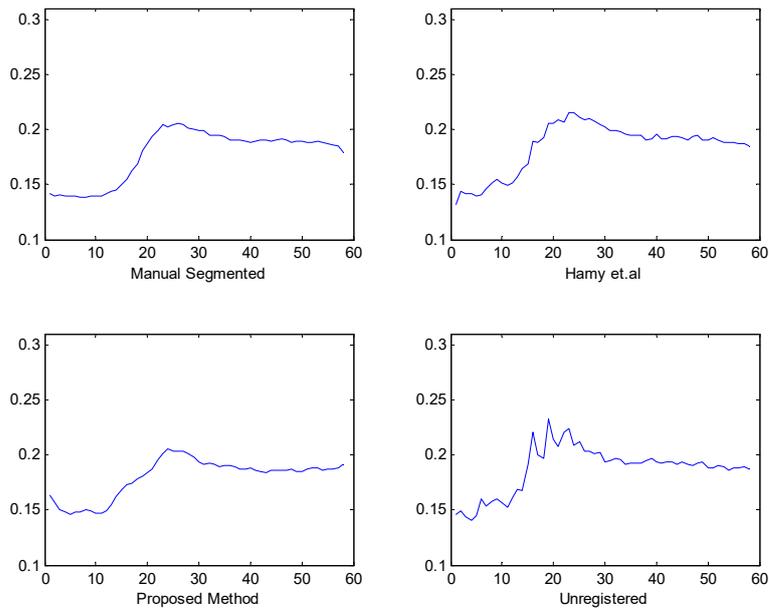

Figure 8: Averaged time-intensity curves of the six sectors defined in the American Heart Association of the myocardium. The proposed method curve is compared with the manual segmented curve, unregistered and the existing method curves. Acceleration rate R=2.

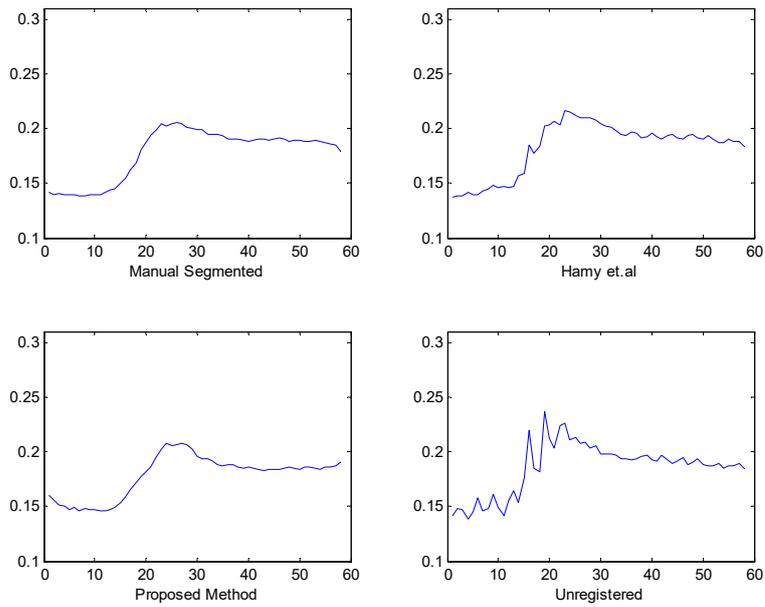

Figure 9: Averaged time-intensity curves of the six sectors defined in the American Heart Association of the myocardium. The proposed method curve is compared with the manual segmented curve, unregistered and the existing method curves. Acceleration rate R=4.

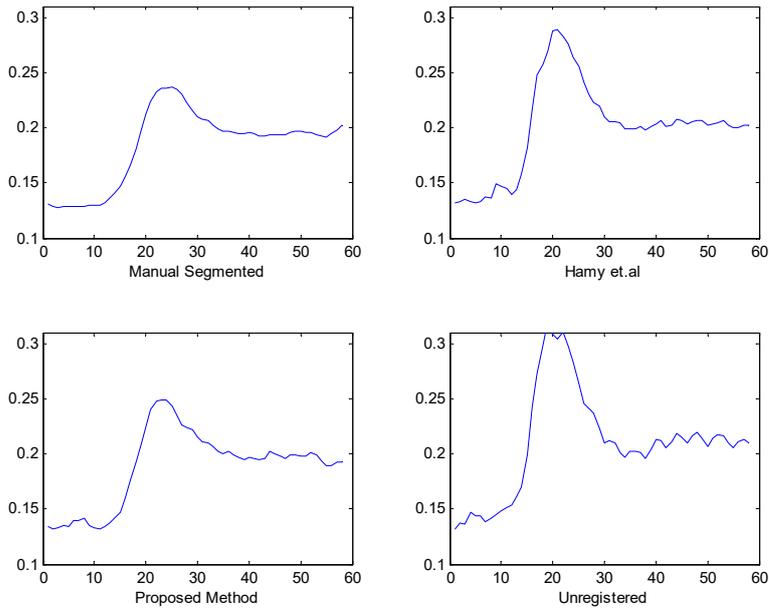

figure 10: Averaged time-intensity curves of the six sectors defined in the American Heart Association of the myocardium. The proposed method curve is compared with the manual segmented curve, unregistered and the existing method curves. Acceleration rate R=8.

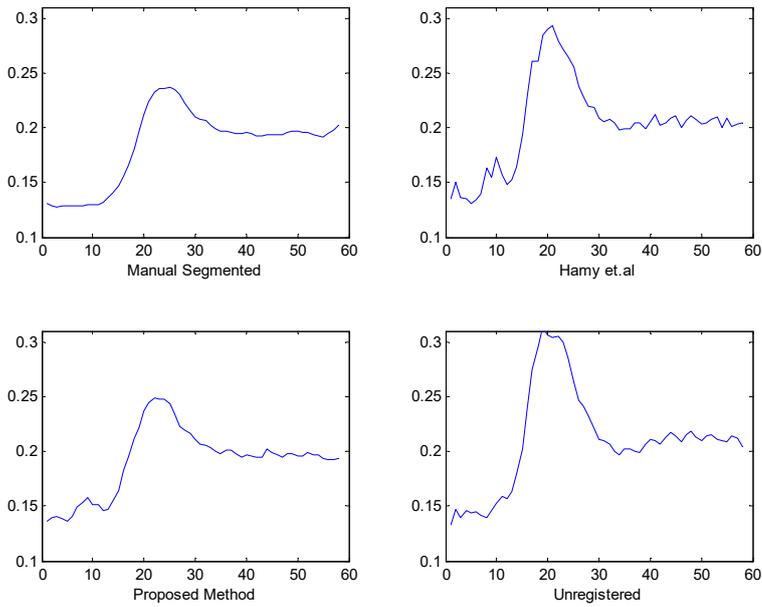

Figure 11: Averaged time-intensity curves of the six sectors defined in the American Heart Association of the myocardium. The proposed method curve is compared with the manual segmented curve, unregistered and the existing method curves. Acceleration rate R=12

## 5. Discussion

In this paper, a novel method was proposed to correct the respiratory motion in cardiac perfusion images. In this work, free breathing undersampled cardiac perfusion data was first reconstructed and decomposed into the low rank component and the sparse component using novel reconstruction method. The sparse component, which constituted the respiratory motion (periodic) and the contrast motion (non-periodic), was further decomposed into the periodic and non-periodic components. The output of this process were the motion free background, the respiratory motion and the contrast motion. Demon based registration method was employed to estimate the motion between the reference respiratory state and other respiratory states to eliminate the respiratory motion artifacts from the cardiac perfusion images.

The performance of the first stage was shown using simulated data and in vivo data with four acceleration rates. Table 1 showed the comparison of the proposed method with the existing L+S method in terms of RMSE using simulated data. Since the SNR=30 was used in the simulated data, therefore, the difference between the methods are nominal at different acceleration rates. Table 2 and Table 3 showed the performance comparison of the proposed method and the L+S method using the in vivo rest perfusion data and the in vivo stress perfusion data respectively. Since the proposed method suppressed the acquisition noise, it performed better in terms of RMSE. Tables also showed that the RMSE deteriorated with increasing acceleration rates.

The second stage in the proposed method separated the respiratory motion from the contrast motion using periodicity decomposition method. The performance of the second stage was assessed using simulated and in vivo images. The output of second stage were the low rank component L, the contrast motion Q, and the respiratory motion P. The first rows in the Figure 1, Figure 2, Figure 3 and Figure 4 showed the combination of the low rank and the contrast component (L+Q) images. This combined perfusion image series can be used to identify the flow of contrast without the presence of respiratory motion and without using the registration algorithms. The only drawback is that the quality of image series are slightly compromised.

In order to improve the image series quality, the combined low rank L and respiratory P image series are registered using Demons based registration method in the third stage. The output of the third stage is the registered perfusion image series without respiratory motion artifacts as shown in the Figure 5, Figure 6, and Figure 7. The performance of the third stage is assessed using simulated images, in vivo perfusion rest images and in vivo perfusion stress images. Figure 5 shows the qualitative performance of the proposed method using under sampled simulated perfusion data. The green horizontal line indicates that the frames are registered to the same respiratory state. In order to analyze the performance of the proposed method, both the rest perfusion images with small respiratory displacement and the stress perfusion images with large respiratory displacement.
Figure 6, and Figure 7 show the registered rest image series and the registered stress image series at different acceleration rates respectively.

Time intensity curves show the comparison of the proposed method with Hamy et. al. motion correction method and the unregistered perfusion images.. Manually segmented time intensity curves are used as the reference curves. The time intensity curves of the manually acquired, proposed method, Hamy et. al. method and the unregistered images are shown in the Figure 8, Figure 9, Figure 10, and Figure 11. Figure 8, and Figure 9 show the time intensity curves of the rest perfusion data at acceleration rates R=2 and R=12, respectively. Figure 10, and Figure 11 depict the time intensity curves of the stress perfusion data at acceleration rates R=2 and R=12, respectively. The proposed method follows the reference curves, whereas the Hamy et. al. method curves illustrates erratic behavior, which is mainly because of the contrast motion with the respiratory motion. The proposed method has shown the promising results, however there are some limitations involved. Optimization and parallelization techniques can be employed to reduce the computational time of the proposed method. The aliasing artifacts due to under sampling in the first stage were observed in the results. The reconstruction algorithm of the first stage can be further improved by employing better reconstruction methods to incorporate the high acceleration rates. Intensity based registration was used in the proposed method, which accumulated the interpolation errors. To reduce the errors, better registration techniques can be used in the proposed method or the motion correction techniques need to be developed without registration methods. The proposed method corrects the motion artifacts in normal patients, however, further improvement may be required to apply to the patients with reduced blood flow and other abnormalities.

## 6. Conclusion

We proposed the robust motion correction technique, which reconstructed the cardiac perfusion images from the highly undersampled data and separated the respiratory motion from the contrast intensity changes. It was successfully tested on the simulated and clinical data in which the proposed technique robustly separated the respiratory motion and corrected the misalignment of the perfusion images. Our technique showed better results as compared to the existing reconstruction and motion correction techniques. The method was validated with manually acquired time intensity curves and it performed better as compared to the unregistered time intensity curves and the existing motion correction method.

**References**

[1]  E. J. Candès, X. Li, Y. Ma, and J. Wright, "Robust principal component analysis?," *Journal of the ACM (JACM),* vol. 58, p. 11, 2011.
[2]  V. Hamy, N. Dikaios, S. Punwani, A. Melbourne, A. Latifoltojar, J. Makanyanga, M. Chouhan, E. Helbren, A. Menys, and S. Taylor, "Respiratory motion correction in dynamic MRI using robust data decomposition registration–Application to DCE-MRI," *Medical image analysis,* vol. 18, pp. 301-313, 2014.
[3]  R. Otazo, E. Candes, and D. K. Sodickson, "Low‐rank plus sparse matrix decomposition for accelerated dynamic MRI with separation of background and dynamic components," *Magnetic resonance in medicine,* vol. 73, pp. 1125-1136, 2015.
[4]  Z. Zhou, X. Li, J. Wright, E. Candes, and Y. Ma, "Stable principal component pursuit," in *2010 IEEE International Symposium on Information Theory*, 2010, pp. 1518-1522.
[5]  M. Tao and X. Yuan, "Recovering low-rank and sparse components of matrices from incomplete and

noisy observations," *SIAM Journal on Optimization,* vol. 21, pp. 57-81, 2011.

[6] M. Li and X. Yuan, "The Augmented Lagrangian Method with Full Jacobian Decomposition and Logarithmic-quadratic Proximal Regularization for Multiple-block Separable Convex Programming."

[7] B. He, H.-K. Xu, and X. Yuan, "On the proximal Jacobian decomposition of ALM for multiple-block separable convex minimization problems and its relationship to ADMM," *Journal of Scientific Computing,* vol. 66, pp. 1204-1217, 2016.

[8] W. A. Sethares and T. W. Staley, "Periodicity transforms," *IEEE transactions on Signal Processing,* vol. 47, pp. 2953-2964, 1999.

[9] J.-P. Thirion, "Image matching as a diffusion process: an analogy with Maxwell's demons," *Medical image analysis,* vol. 2, pp. 243-260, 1998.

[10] L. Wissmann, C. Santelli, W. P. Segars, and S. Kozerke, "MRXCAT: Realistic numerical phantoms for cardiovascular magnetic resonance," *Journal of Cardiovascular Magnetic Resonance,* vol. 16, p. 1, 2014.

[11] G. Wollny and P. Kellman, "Free breathing myocardial perfusion data sets for performance analysis of motion compensation algorithms," *GigaScience,* vol. 3, p. 1, 2014.

[12] A. H. Ahmed, I. M. Qureshi, J. A. Shah, and M. Zaheer, "Motion correction based reconstruction method for compressively sampled cardiac MR Imaging," *Magnetic Resonance Imaging,* 2016.